\documentclass{svjour2}                    
\smartqed  
\usepackage{graphicx}
\usepackage{amsmath}
\usepackage{enumerate}
\journalname{Journal of Statistical Physics}
\begin{document}

\title{Traces of Integrability in Relaxation of One-Dimensional Two-Mass Mixtures
}


\author{Zaijong Hwang \and Frank Cao \and Maxim Olshanii}


\institute{Zaijong Hwang \at Department of Physics, University of Massachusetts Boston\\
\email{zaijong.hwang001@umb.edu}
\and Frank Cao \at Department of Physics, University of Connecticut\\
\email{cao@phys.uconn.edu}
\and Maxim Olshanii \at Department of Physics, University of Massachusetts Boston\\
\email{maxim.olchanyi@umb.edu}
}

\date{Received: date / Accepted: date}

\maketitle

\begin{abstract}
We study relaxation in a one-dimensional two-mass mixture of hard-core particles. A heavy-light-heavy triplet of three neighboring particles can form a little known unequal mass generalization of Newton's cradle at particular light-to-heavy mass ratios. An anomalous slow-down in the relaxation of the whole system is expected due to the presence of these triplets, and we provide numerical evidence to support this prediction. The expected experimental realization of our model involves mixtures of two internal states in optical lattices, where the ratio between effective masses can be controlled at will.

\end{abstract}

\section{Introduction}

A large finite collection of equal mass, hard-core, point particles undergoing elastic collisions in one dimension with periodic boundary conditions is a strictly non-ergodic system \cite{tonks1936}. Ergodicity is expected to emerge however, should the masses be made unequal \cite{redner2006}. Yet while such a system of unequal mass particles may be ergodic as a whole---any observable of the whole system is expected to relax towards its ensemble average---collisions between particles within a subset of the system may not always contribute to the overall relaxation of the whole system. This of course is trivially true for subsets that only consist of equal mass particles, but we will demonstrate that it is also true for certain subsets of three unequal mass particles that form a generalized Newton's cradle. 

\section{A Generalization of Newton's Cradle with Three Particles}

Newton's cradle is usually encountered as a simple device used to demonstrate the conservation of energy and momentum. The typical cradle consists of a line of identical pendulums. When a pendulum on one end is lifted and released, it will fall and strike the other stationary pendulums with some velocity. After a cascade of two-body collisions, the pendulum on the other end will be ejected with that same velocity, while all other pendulums are now stationary. There are numerous variations to this basic process, such as imparting multiple pendulums with an initial velocity before the cascade. Although there are many subtleties underlying the operation of the physical Newton's cradle \cite{schmalzle1980}, we will overlook these and only focus on one particular phenomenon: if equal mass particles are indistinguishable, then the set of mass-velocity pairs before and after a cascade of collisions should be identical in a Newton's cradle. Taking this as a defining characteristic, we will generalize Newton's cradle to include unequal masses, while still conserving the mass-velocity pairs over cascades of collisions.

Consider a system of three hard-core particles on an open line. Let their masses be $m_1$, $m_2$, $m_3$ with positions $x_1$, $x_2$, $x_3$. The Lagrangian of this system may be written as
	\begin{align}
	L &= \frac{1}{2}\sum\limits_{i=1}^3 m_i \dot{x}_i^2 - V(x_1, x_2, x_3).
	\end{align}
Without loss of generality, we may choose $x_1 \leq x_2 \leq x_3$, which will fix the potential as
	\begin{align}
	V(x_1, x_2, x_3) &= \left\{
		\begin{array}{ll}
		 	0 & \quad \text{if } x_1 \leq x_2 \leq x_3\\
  		\infty & \quad \text{otherwise}
		\end{array}
	\right..
	\end{align}
Using a set of coordinates that produces a scalar mass for the kinetic energy of the relative motion of three particles \cite{hirschfelder1959,mcguire1964}
	\begin{align}
	x &= \sqrt{\frac{m_1 m_2}{m_1+m_2}}\left(x_1-x_2\right),\\
	y &= \sqrt{\frac{m_3\left(m_1+m_2\right)}{m_1+m_2+m_3}}\left(\frac{m_1 x_1+m_2 x_2}{m_1+m_2}-x_3\right),\\
	z &= \frac{1}{\sqrt{m_1+m_2+m_3}}\left(m_1 x_1+m_2 x_2+m_3 x_3\right),
	\end{align}
the Lagrangian can be transformed into
	\begin{align}
	L &= \frac{1}{2}\left(\dot{x}^2 + \dot{y}^2 + \dot{z}^2 \right) + V(x, y, z),
	\end{align}
with a potential of
	\begin{align}
	V(x, y, z) &= \left\{
		\begin{array}{ll}
	 		0 & \quad
	 			\displaystyle{\text{if } x \leq 0 \text{ and }
	 				y \leq \left(\frac{m_1 m_3}{m_2\left(m_1+m_2+m_3\right)}\right)^\frac{1}{2} x
	 			}\\
  		\infty & \quad \text{otherwise}
		\end{array}
	\right..
	\end{align}
The coordinates above are very similar to the standard Jacobi coordinates that are frequently used in solving many-body problems, but the prefactors are substantially different here. The transformation above will always produce a scalar mass, whereas Jacobi coordinates can only guarantee a diagonal mass matrix. The importance of this distinction will become evident shortly.

\begin{figure}
\centering
\includegraphics[scale=0.5]{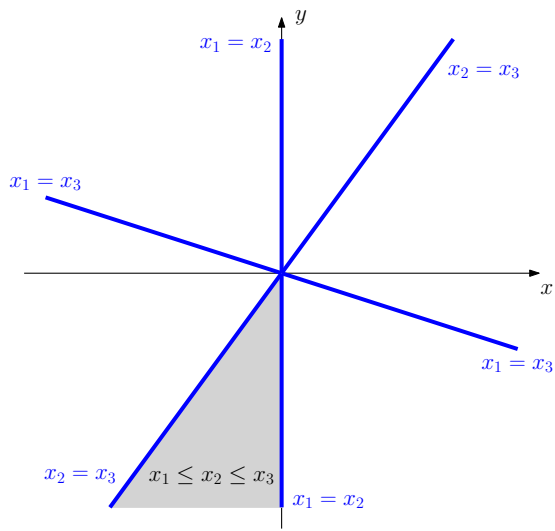}
\caption{The $x$ and $y$ transformed coordinates for the relative motion of three particles on a line. The system is confined to the $x_1\leq x_2\leq x_3$ shaded domain by choice. There are six wedged regions that correspond to the possible permutations of three particles on a line. In this figure, we have set $m_1=m_3=M$ and $m_2=m$ with $m/M = \sqrt{5}-2$. The angle between the $x_1=x_2$ and $x_2=x_3$ line here is $\pi/5$.}
\label{fig:jacobi_coordinates}
\end{figure}

The new $z$ coordinate is proportional to the center of mass coordinate of the three particles. Since the time evolution of the center of mass is decoupled from the collision dynamics of the particles, we may eliminate the $z$ coordinate from our analysis. The motion of three particles in one dimension can thus be mapped into the motion of one particle of unit mass in two dimensions as seen in Fig.\nobreak\ \ref{fig:jacobi_coordinates}.

On this two-dimensional plane, the relative positions of the three particles are encoded within the $x$ and $y$ coordinates. The trajectory of the three particles is represented by a ray, the orientation of which determines the relative velocities of the three particles up to a rescaling of time. As a consequence of choosing $x_1\leq x_2\leq x_3$, the single point that now represents the relative motion of three particles will be constrained to move within a wedged region bounded by two hard walls.

The scalar mass of this point now becomes important as specular reflection is guaranteed to occur whenever the point impacts these walls. The overall setup here resembles that of a dynamical billiard, except for the open boundaries. A proper triangular billiard in this two-dimensional plane with closed boundaries would correspond to a system of two particles in one dimension bounded by two hard walls \cite{glashow1997,guarneri1997,prosen2014}. 

\begin{figure}
\centering
\includegraphics[scale=0.5]{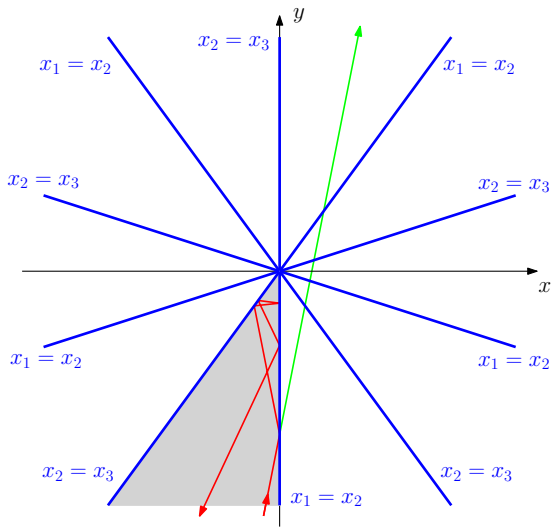}
\caption{Applying the method of images when the light-to-heavy mass ratio is $m/M = \sqrt{5}-2$ results in $\theta=\pi/5$. As this corresponds to an odd kaleidoscope, the incoming rays will not in general be parallel to the outgoing rays as in the case of even kaleidoscopes. Shaded region is the $x_1\leq x_2\leq x_3$ domain as seen in Fig.\nobreak\ \ref{fig:jacobi_coordinates} and now acts as a fundamental chamber of the corresponding reflection group.}
\label{fig:method_of_images}
\end{figure}

The presence of specular reflection allows our transformed system to be studied with the method of images as in Fig.\nobreak\ \ref{fig:method_of_images}. Consider the case where the $x_1=x_2$ and $x_2=x_3$ boundaries cross at an angle of $\theta=\pi/n$ where $n$ is an integer. This allows the wedged region in which the system is confined within to tile the $x$-$y$ plane through reflections across lines of contact between the particles.

This tiling is significant because it ensures that a given set of initial velocities $v_1$, $v_2$, $v_3$ for three particles will always produce a unique set of final velocities $v_1'$, $v_2'$, $v_3'$ regardless of the order of collisions between the particles. In terms of the rays that represent the trajectory of the system, tiling allows a set of parallel incoming rays to remain parallel with each other after they undergo a cascade of reflections, no matter which wall the rays first connect with. This property can be deduced from Fig.\nobreak\ \ref{fig:method_of_images} by tracing the trajectory of the green line in reverse: such a path will reveal the set of reflections that would occur should the incoming ray have connected with the $x_2=x_3$ boundary first.

There is a rich mathematical background underlying the system depicted in Fig.\nobreak\ \ref{fig:method_of_images}, and we shall make a short digression to introduce some relevant concepts. The shaded region in the figure is the fundamental chamber where the physical system actually resides. The two boundaries enclosing the fundamental chamber can be treated as mirrors, and together they form a kaleidoscope. The normal vector of a mirror is called a root vector, and the two root vectors of a kaleidoscope can be used to generate a collection of root vectors that is closed under reflections by the corresponding mirrors. This collection of root vectors forms a root system, which is in turn associated with a reflection group\footnote{We are omitting some details regarding the length of root vectors, which can influence whether the relevant kaleidoscope corresponds to a physical system of particles that are on a line, ring, or confined to an interval.}. The kaleidoscope in Fig.\nobreak\ \ref{fig:method_of_images} corresponds to the root system $I_2(5)$ and is associated with the dihedral group of the decagon. In general, two-dimensional kaleidoscopes with mirrors at an angle of $\theta=\pi/n$ will correspond to the $I_2(n)$ root systems and be associated with the symmetries of the regular $2n$-gon \cite{coxeter_book_regular_polytopes}.

Returning to the two-mass mixture that is the subject of this study, we shall set $m_1=m_3=M$ to be the heavy mass and $m_2=m$ to be the light mass. The mass ratio can then be related to the angle $\theta=\pi/n$ as
	\begin{align}
	\frac{m}{M} &= \frac{1-\cos\left(\pi/n\right)}{\cos\left(\pi/n\right)}.
	\label{mass_ratios}
	\end{align}

If $n$ is odd, a system point which has not experienced any previous collisions will undergo a cascade of $n$ collisions with the walls. The angle between the incoming ray and the first wall will also be equal to that between the outgoing ray and the last wall. In the center of mass frame, this results in the exchange of velocities between the left and right heavy particles while the middle light particle retains its velocity. The mass-velocity pairs will be conserved even if the center of mass velocity is not zero, and such a triplet will therefore constitutes a generalized Newton's cradle. The actual outcome of a cascade at odd $n$ is
	\begin{align}
	\left\{
		\begin{array}{cc}
		v_1\\
		v_2\\
		v_3\\
		\end{array}
	\right\}
	\xrightarrow[\text{odd cascade}]{\text{$n$ collisions}}
	\left\{
		\begin{array}{cc}
		v_3\\
		v_2\\
		v_1\\
		\end{array}
	\right\}.
	\end{align}

Imagine now that three particles with odd $n$ are embedded within a gas of some other particles. Most of the time, this triplet will not manifest the behavior of a Newton's cradle, because the necessary cascade of $n$ two-body collisions are likely to be interrupted by collisions with neighboring particles of the host gas. With finite probability however, the three particles may find themselves in such close proximity that they can complete a cascade of $n$ two-body collisions {\it uninterrupted}. Given such a complete revival of the local velocity distribution, one would expect a slowdown in the relaxation rate of the many-body system as a whole. This effect will be our focus in the next section.

If $n$ is even, a system point which has not experienced any previous collisions will also undergo a cascade $n$ collisions with the walls, but the incoming and outgoing rays will now remain parallel to each other. In the center of mass frame, this corresponds to a complete reversal of all particle velocities after such a cascade of collisions. The resulting change in velocity distribution is
	\begin{align}
	\left\{
		\begin{array}{cc}
		v_1\\
		v_2\\
		v_3\\
		\end{array}
	\right\}
	\xrightarrow[\text{even cascade}]{\text{$n$ collisions}}
	\left\{
		\begin{array}{cc}
		V_{CM} - \left(v_1 - V_{CM}\right)\\
		V_{CM} - \left(v_2 - V_{CM}\right)\\
		V_{CM} - \left(v_3 - V_{CM}\right)
		\end{array}
	\right\},
	\end{align}
where $V_{CM}$ is the center of mass velocity. In principle, such a process is not expected to change the relaxation rate of the many-body system as a whole, although there exist conditions in which it will come into effect. These circumstances will be relevant in the next section as well.

\section{Numerical Simulation of a Two-Mass Mixture}

\begin{figure}
\centering
\includegraphics[scale=0.6]{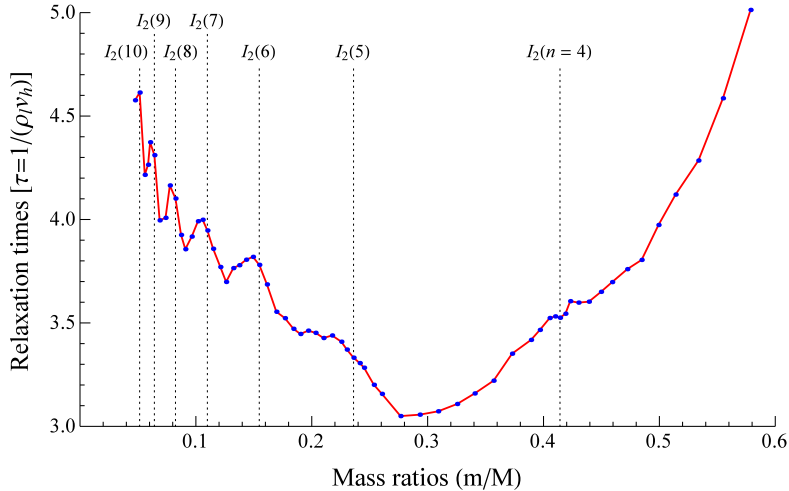}
\caption{Relaxation time of $\chi\equiv\left< E_i^2 \right>/\left<E_i\right>^2$ (see main text for full definition) as a function of mass ratio. Data points are in blue, and joined by red lines to aid visualization. The vertical axis is the relaxation time in units of $1/\left(\rho_l v_h\right)$, where $\rho_l$ is the number density of the light particles, and $v_h$ is the initial speed of the heavy particles. This unit of measurement is an order-of-magnitude estimate for a given heavy particle's initial mean free time between collisions with light particles. Vertical dotted lines correspond to mass ratios where $\theta=\pi/n$ for integer $n$ (see equation \ref{mass_ratios}) and peaks are expected to occur. These lines are labeled by the associated $I_2(n)$ root system, starting from $\theta=\pi/10$ on the left and progressing to $\theta=\pi/4$ on the right. The rising background likely contributes to the slight displacement of the peaks from their expected positions.}
\label{fig:relaxation_times}
\end{figure}

The slowdown in the relaxation rate for a large system due to the presence of heavy-light-heavy triplets that form a generalized Newton's cradle is demonstrated by a numerical simulation with the results shown in Fig.\nobreak\ \ref{fig:relaxation_times}. The initial condition consists of 100 light and 200 heavy particles randomly positioned within the $[0,1]$ interval with periodic boundary conditions. We set $M=1$ and adjust $m$ to achieve different mass ratios. All light particles are initially stationary, while half the heavy particles are randomly chosen to have initial momentum $+1$, and the other half $-1$.

In order to determine the rate of thermalization, we need a suitable measure that will place light and heavy particles on an equal footing. The distribution of kinetic energies should then be a good indicator of thermalization due to the eventual equipartition between light and heavy particles. To measure how fast the kinetic energy approaches the Boltzmann distribution, we choose to examine the first non-trivial moment of the kinetic energy distribution, standardized by the mean kinetic energy, as follows
	\begin{align}
	\chi&\equiv\frac{\left< E_i^2 \right>}{\left<E_i\right>^2},
	\end{align}
where the averages are taken over all $i=1\ldots 300$ particles. It will have a value of 1.5 under our initial conditions and is expected to rise to 3 if the system is fully thermalized to a Boltzmann distribution. Since equal mass collisions do not contribute to relaxation, the simulation is carried out over a fixed length of 10,000 light-heavy collisions. At each mass ratio, the simulation is repeated 5,000 times and the average value of $\chi$ as a function of the number of light-heavy collisions over these realizations is fit to an exponential curve to extract a time constant.

We choose to use the mean initial frequency of light-heavy collisions as a unit of measurement for the rate of thermalization, because light-heavy collisions are the fundamental interactions that drive thermalization: we expect them to be involved in any universal results that might emerge in the relaxation of a two-mass mixture. One prominent universal result in this vein \cite{wieman1993,foot1996} is that the kinetic energy of hard-core spheres in three dimensions will thermalize across dimensions in 2.7 collisions per particle over a broad range of initial conditions.

Figure\nobreak\ \ref{fig:relaxation_times} demonstrates that there are clear peaks in the relaxation time at the expected mass ratios where $\pi/\theta$ is an integer. The background curve also exhibits two expected features. As $m/M\to 1$ on the right, the relaxation time should diverge as the system is approaching the exactly integrable equal mass configuration. When $m/M\to 0$ on the left, the relaxation time should also diverge since it will take infinitely many light-heavy collisions to change the velocity of the heavy particles.

Notice that the peaks are present at {\it both} odd and even values of $n$. As discussed in the previous section, peaks in relaxation time are expected around odd values of $n$ as a result of local triplets forming a generalized Newton's cradle. For even values of $n$, a complete uninterrupted cascade of two-body collisions merely reverses all velocities in the center of mass frame, and there is no {\it a priori} expectation that this will result in a slowdown in the overall rate of relaxation for the many-body system as a whole. Two effects may contribute to these peaks at even $n$:

\begin{enumerate}
\item The tiling property for all kaleidoscopes guarantees that any set of initial velocities will always produce a unique set of final velocities after a complete cascade of collisions. The uncertainty in the final velocity distribution is therefore lower at these kaleidoscopic mass ratios when compared to a generic sequence of collisions, for both even and odd $n$.
\item  With our initial velocity distribution, any heavy-light-heavy triplet that is ready to undergo a complete cascade will consist of two heavy particles approaching each other with opposite velocities, and light partile at rest in between them. For this initial condition, the reflection of all velocities about the center of mass velocity after a complete cascade is equivalent to a revival of the initial velocity distribution. As the many-body system continues to relax however, memory of the initial distribution should also erode and weaken this effect.
\end{enumerate}

Neither effects above have been quantified in detail, and a complete explanation for the increase in relaxation time at mass ratios with even $n$ is still required.

\section{Conclusions and Outlook}

Using simple geometric arguments, we showed that a triplet of heavy-light-heavy hard-core particles in one dimension can form a generalized Newton's cradle at particular light-to-heavy mass ratios. This generalized Newton's cradle will preserve the velocity distribution of its constituent particles as they undergo a cascade of collisions: a series of two-body collisions that begins with the triplets not having any prior collisions, and ends with them moving away from each other and never colliding again. We provided numerical evidence that the presence of this generalized Newton's cradle at the three-body level manifests itself in a larger many-body system as peaks in the relaxation time at particular mass ratios.

The expected experimental realization of our model would involve a mixture of two internal states of a single atomic species in an optical lattices, where the ratio between the lattice induced effective masses can be controlled at will. Previous experiments have already produced a mixture of two superfluids of comparable controllable masses \cite{schneble2010}, and a mixture where one internal state has effectively infinite mass \cite{schneble2011}.

The present work is a natural continuation of the quest initiated by Michel Gaudin \cite{gaudin2014_book,gaudin1971} to connect exactly solvable particle problems with kaleidoscopes and root systems. Gaudin originally suggested that only two types of root systems would be relevant to particle problems: $A_{N-1}$ for $N$ equal mass particles on a ring, and $C_N$ for $N$ equal mass particles between two hard walls on a line. He further demonstrated that any particle problem treatable as a kaleidoscope could be quantized and solved by Bethe Ansatz \cite{gaudin2014_book,emsiz2006,emsiz2009,emsiz2010}. 

By considering unequal mass particles, we extend the set of root systems that are relevant to particle problems to include $I_2(n)$ for three unequal mass particles on a line with mass ratios prescribed by equation (\ref{mass_ratios}). We expect quantized three particle problems with kaleidoscopic mass ratios will also be solvable by Bethe Ansatz. A more general theory for the connection between particle problems and root systems---including an exactly solvable 4-body problem in a box associated with the $F_4$ root system---will be provided elsewhere \cite{olshanii2015}.

There are several questions that remain open for study. Both the peak-to-background ratio and overall background in Fig.\nobreak\ \ref{fig:relaxation_times} require a quantitative theoretical foundation. A better explanation for the peaks in relaxation time at even $n$ is also needed. An interesting possibility for future research is to further extend the set of exactly solvable particle problems by linking previously unexplored root systems to novel particle problems, including those with unequal masses \cite{olshanii2015}.



\begin{acknowledgements}
We thank S. Jackson, S. Redner, V. Dunjko, D. Schneble, J.-S. Caux, and J. McGuire for useful discussions. This work was supported by grants from the Office of Naval Research ({\it N00014-12-1-0400}) and the National Science Foundation ({\it PHY-1402249}).

\end{acknowledgements}


\bibliographystyle{spphys}
\bibliography{1D_Collisions}

%
%

\end{document}